\documentclass[aps,reprint,groupedaddress]{revtex4-1}
\usepackage{amsmath}
\usepackage{bm} % bold math
\usepackage{amssymb}
\usepackage{color}
\usepackage{multirow}
\usepackage[colorlinks,bookmarks=false,citecolor=red,linkcolor=blue,urlcolor=blue]{hyperref}
\usepackage{graphicx}
\usepackage{mathrsfs,bbm}
\usepackage[caption=false,subcaption]{subfig} \usepackage{wrapfig}
\usepackage[version=3]{mhchem}
\usepackage{float}
\usepackage{dcolumn}
\usepackage{calligra}
\DeclareMathAlphabet{\mathcalligra}{T1}{calligra}{m}{n}
\DeclareFontShape{T1}{calligra}{m}{n}{<->s*[2.2]callig15}{}

\def\A{{\bf A}}

\def\Q{{\bf Q}}

\def\k{{\bf k}}
\def\r{{\bf r}}

\def\dr{{\bf dr}}
\def\bmdelta{\bm{\delta}}
\def\bmsigma{\bm{\sigma}}
\def\bmtau{\bm{\tau}}
\def\bmdelta{{\bm \delta}}

\def\sbar{\bar{s}}
\def\sfz{\mathsf{z}}
\def\Hhat{\hat{H}}

\def\ahat{\hat{a}}
\def\bhat{\hat{b}}
\def\chat{\hat{c}}
\def\fhat{\hat{f}}
\def\nhat{\hat{n}}
\def\shat{\hat{s}}
\def\that{\hat{t}}
\def\xhat{\hat{x}}

\def\zhat{\hat{z}}

\def\phihat{\hat{\phi}}
\def\psihat{\hat{\psi}}

\def\nhat{\hat{n}}
\def\fhat{\hat{f}}

\def\ttilde{\tilde{t}}
\def\phitil{\tilde{\phi}}
\def\psitil{\tilde{\psi}}

\begin{document}
\title{Inversion and magnetic quantum oscillations in symmetric periodic Anderson model}
\author{Panch Ram} 
\author{Brijesh Kumar}
\email{bkumar@mail.jnu.ac.in}
\affiliation{School of Physical Sciences, Jawaharlal Nehru University, New Delhi 110067, India.}
\date{\today}  
%%%%%%%%%%%%%%%%%%% 
\begin{abstract}
We study the symmetric periodic Anderson model of the conduction electrons hybridized with the localized correlated electrons on square lattice. Using the canonical representation of electrons by Kumar, we do a self-consistent theory of its effective charge and spin dynamics, which produces an insulating ground state that undergoes continuous transition from the Kondo singlet to N\'eel antiferromagnetic phase with decreasing hybridization, and uncovers two inversion transitions for the charge quasiparticles. With suitably inverted quasiparticle bands for moderate to weaker effective Kondo couplings, this effective charge dynamics in the magnetic field coupled to electronic motion produces magnetic quantum oscillations with frequency corresponding to the half Brillouin zone. 
\end{abstract}
%\pacs{75.10.Jm, 75.10.Kt, 75.30.Kz, 05.30.Rt}
\maketitle

%%%%%%%%%%%%%%%%%%%%%%%%%%%%%%

\section{Introduction \label{sec:intro}}
The magnetic quantum oscillations periodic in inverse magnetic field, namely the de Haas-van Alphen (dHvA) effect~\cite{Haas1934}, were long held to be an exclusive property of the metals, and have provided the means to measure the Fermi surfaces thereof (as noted by Onsager~\cite{Onsager}). The recent discovery of dHvA oscillations in \ce{SmB6}, a Kondo insulator, challenges this conventional view~\cite{Li2014,Tan2015}. It has forced us to reexamine the physics of Kondo insulators, and reconsider the dHvA effect. The origin of dHvA oscillations in insulators (Kondo or otherwise) is a vigorously pursued problem, with a growing number of theoretical studies and some interesting proposals~\cite{Kishigi2014,Knolle2015,Zhang2016,Erten2016,Pal2016,Ram2017,Sodemann2018,Shen2018,Harrison2018}.

The Kondo insulators are a class of heavy-fermion systems~\cite{Fisk1996,Coleman2015}, of which the \ce{SmB6}~\cite{Menth1969}, \ce{YbB12}~\cite{Iga1988}, \ce{Ce3Bi4Pt3}~\cite{Hundley1990} are some of the best known examples. They behave as paramagnetic insulators (with small gap) at sufficiently low temperatures. But at high temperatures, they are Curie-Weiss metals. Recently, the quantum oscillations have been reported to occur also in \ce{YbB12}~\cite{Liu2018}. The basic physical setting of a Kondo insulator involves the localized $f$ electrons in hybridization with the conduction electrons. The minimal model that applies to the heavy-fermion systems is the periodic Anderson model (PAM) with local repulsion, $U$, for the $f$ electrons and the hybridization, $V$, between the $f$ and the conduction electrons~\cite{Anderson1961,Hewson1993,Coleman2015}. At half-filling, the PAM describes the Kondo insulators. 

To exactly realize the half-filling, it is common to consider the particle-hole symmetric PAM with nearest-neighbor conduction electron hopping, $t$, on bipartite lattices~\cite{Ueda1992,Jarrell1995,Vekic1995,Smith2003}. The Hamiltonian of the {\em symmetric} periodic Anderson model (SPAM) can be written as:
\begin{align}
\Hhat =&-t\sum_{\r,\bmdelta}\sum_{s=\uparrow,\downarrow}\chat^\dag_{\r,s} \chat^{ }_{\r+\bmdelta,s}
-V\sum_\r\sum_{s=\uparrow,\downarrow} \left[\chat^\dag_{\r,s} \fhat^{ }_{\r,s} 
+{\rm h.c.} \right] \nonumber \\
&+U \sum_\r \left(\nhat^{f}_{\r,\uparrow}-\frac{1}{2}\right)\left(\nhat^{f}_{\r,\downarrow}-\frac{1}{2}\right) 
\label{eq:SPAM}
\end{align}
where $\r$ is summed over the lattice sites, $\bmdelta$ denotes the nearest-neighbors of $\r$, the fermion operators $\chat^{\dag}_{\r,s}$ ($\chat^{ }_{\r,s}$) create (annihilate) the conduction electrons, $\fhat^{\dag}_{\r,s}$ ($\fhat^{ }_{\r,s}$) do likewise for the localized electrons, and $ \nhat^{f}_{\r,s}=\fhat^{\dag}_{\r,s} \fhat^{ }_{\r,s}$.  

In continuation of our recent work on magnetic quantum oscillations in Kondo insulators~\cite{Ram2017}, we in this paper investigate the dHvA oscillations in the ground state of the SPAM. The theory of Kondo insulators, as developed by us for the Kondo lattice model (KLM) in Ref.~\onlinecite{Ram2017}, is applied here to study the orbital response of the SPAM to a uniform magnetic field. Notably, with this theory of SPAM in the representation of electrons by Kumar~\cite{BK2008}, we discover two inversion transitions, one each for the charge quasiparticles with narrow and broad dispersions, as $V$ decreases for a given $U$. As for the KLM, here too, we find the quasiparticle band inversion to be the key determinant for the dHvA oscillations to occur or not to occur. Hence, for the strong Kondo couplings ($\sim V^2/U$), we see no dHvA oscillations. But in the regime of intermediate to weaker Kondo couplings, where the quasiparticle bands are appropriately inverted, we obtain the dHvA oscillations in the insulating ground state of the SPAM both in the Kondo singlet and the ordered antiferromagnetic (AFM) phases. Notably, the magnetic oscillations from the two kinds of quasiparticles are found to be mutually out of phase, due to which a non-trivial partial cancellation occurs. But still the net magnetization prominently oscillates with a frequency that corresponds to the half of the Brillouin zone (BZ).

This paper is organized as follows. In Sec.~\ref{sec:theory}, we formulate a self-consistent theory of the spin and charge dynamics of the SPAM. Using this theory, we describe in Sec.~\ref{sec:ground-state-results} the magnetic quantum phase transition from the Kondo singlet to the N\'eel AFM phase. We then describe the properties of the charge quasiparticles in Sec.~\ref{sec:2inversions}. In particular, there we discuss the inversion of the charge quasiparticle dispersions with decreasing $V$, for a fixed $U$. Through this effective charge dynamics, in the Peierls-coupled uniform magnetic field, we investigate and obtain the dHvA oscillations in the insulating ground state of the SPAM in Sec.~\ref{sec:QOs}. We conclude in Sec.~\ref{sec:conclusion} with a summary of our key findings.

%%%%%%%%%%%%%
\section{Self-consistent theory of charge and spin dynamics \label{sec:theory}}
To study the properties of the SPAM, we use the following canonical representation for $c$ (conduction) and $f$ (localized) electron operators on the two sublattices ($\mathcal{A}$ and $\mathcal{B}$) of a bipartite lattice, as prescribed in Ref.~\onlinecite{BK2008}. While the present consideration applies to any bipartite lattice, but later in this paper, we will work only on the square lattice.
\begin{equation}
\begin{array}{|l|l|}
\hline \r \in \mathcal{A}~{\rm sublattice} ~&~ \r \in \mathcal{B}~{\rm sublattice} \\ \hline ~&~ \\
\chat^\dag_{\r\uparrow} = \phihat_{a,\r}\sigma^+_\r ~&~ \chat^\dag_{\r\uparrow} = i\psihat_{b,\r}\sigma^+_\r \\
 \chat^\dag_{\r\downarrow} = \frac{1}{2}(i\psihat_{a,\r} - \phihat_{a,\r}\sigma^z_\r) ~&~ \chat^\dag_{\r\downarrow} = \frac{1}{2}(\phihat_{b,\r}-i\psihat_{b,\r}\sigma^z_\r) \\ \hline ~&~ \\
\fhat^\dag_{\r\uparrow} = i\psitil_{a,\r}\tau^+_\r ~&~ \fhat^\dag_{\r\uparrow} = \phitil_{b,\r}\tau^+_\r \\
\fhat^\dag_{\r\downarrow} = \frac{1}{2}(\phitil_{a,\r}-i\psitil_{a,\r}\tau^z_\r) ~&~ \fhat^\dag_{\r\downarrow} = \frac{1}{2}(i\psitil_{b,\r} - \phitil_{b,\r}\tau^z_\r)\\ \hline
\end{array}
\label{eq:cf-kumar}
\end{equation}
In Eq.~\eqref{eq:cf-kumar}, $\phihat_{a,\r} = \ahat^\dag_{c,\r}+\ahat^{ }_{c,\r}$, $i\psihat_{a,\r} = \ahat^\dag_{c,\r}-\ahat^{ }_{c,\r}$ and $\phitil_{a,\r} = \ahat^\dag_{f,\r}+\ahat^{ }_{f,\r}$, $i\psitil_{a,\r} = \ahat^\dag_{f,\r}-\ahat^{ }_{f,\r}$ are the Majorana operators corresponding to the spinless fermions $\ahat_{c,\r}$ and $\ahat_{f,\r}$ on the $\mathcal{A}$ sublattice. Similarly, $\phihat_{b,\r} = \bhat^\dag_{c,\r}+\bhat^{ }_{c,\r}$, $i\psihat_{b,\r} = \bhat^\dag_{c,\r}-\bhat^{ }_{c,\r}$ and $\phitil_{b,\r} = \bhat^\dag_{f,\r}+\bhat^{ }_{f,\r}$, $i\psitil_{b,\r} = \bhat^\dag_{f,\r}-\bhat^{ }_{f,\r}$ are the Majorana operators corresponding to the spinless fermions $\bhat_{c,\r}$ and $\bhat_{f,\r}$ on the $\mathcal{B}$ sublattice. The $\sigma_{\r}^{\pm},~\sigma_{\r}^{z}$ and $\tau_{\r}^{\pm},~\tau_{\r}^{z}$ are the Pauli operators. Here, the spinless fermions describe the charge fluctuations, and the Pauli operators describe the electronic spin (or pseudo-spin).

In this representation, the SPAM given by Eq.~(\ref{eq:SPAM}) on a bipartite lattice reads as:
\begin{align}
\Hhat =& -\frac{it}{2} \sum_{\r\in \mathcal{A}} \sum_\bmdelta \left[ \psihat_{a,\r}\phihat_{b,\r+\bmdelta} + \psihat_{b,\r+\bmdelta}\phihat_{a,\r} \left(\bmsigma_\r\cdot\bmsigma_{\r+\bmdelta}\right)\right] \nonumber \\
&  -\frac{iV}{2} \sum_{\r\in \mathcal{A}} \left[\psihat_{a,\r}\phitil_{a,\r} + \psitil_{a,\r}\phihat_{a,\r} \left(\bmsigma_\r\cdot\bmtau_{\r}\right)\right] \nonumber \\
& -\frac{iV}{2} \sum_{\r\in \mathcal{B}} \Big[\psitil_{b,\r}\phihat_{b,\r} 
+ \psihat_{b,\r}\phitil_{b,\r} \left(\bmsigma_\r\cdot\bmtau_{\r}\right)\Big] \nonumber \\
&-\frac{U}{2}\left[\sum_{\r\in \mathcal{A}} \ahat_{f,\r}^{\dag}\ahat_{f,\r} + 
\sum_{\r\in \mathcal{B}} \bhat_{f,\r}^{\dag}\bhat_{f,\r}\right]+ \frac{U L}{4}.
\label{eq:SPAMKumar}
\end{align}
Following Ref.~\cite{Ram2017}, we decouple the spinless fermions from the Pauli operator terms in Eq.~\eqref{eq:SPAMKumar}. In this approximation, the SPAM reads as: $\Hhat \approx \Hhat_c+\Hhat_s+e_0 L$, where $e_0=-(\sfz t\zeta_1\rho_1+2V\zeta_2\rho_0)/4$, 
\begin{align}
\Hhat_{c} =& -\frac{it}{2} \sum_{\r\in \mathcal{A}} \sum_\bmdelta \left[ \psihat_{a,\r}\phihat_{b,\r+\bmdelta} + \rho_1 \psihat_{b,\r+\bmdelta}\phihat_{a,\r}\right] \nonumber \\
& -\frac{iV}{2} \sum_{\r\in \mathcal{A}} \left[\psihat_{a,\r}\phitil_{a,\r} + \rho_0~\psitil_{a,\r}\phihat_{a,\r}\right] \nonumber \\
&-\frac{iV}{2} \sum_{\r\in \mathcal{B}} \biggl[\psitil_{b,\r}\phihat_{b,\r} + \rho_0\psihat_{b,\r}\phitil_{b,\r}\biggr] \nonumber \\
& - \frac{U}{2}\left[\sum_{\r\in \mathcal{A}} \ahat_{f,\r}^{\dag}\ahat_{f,\r} + 
\sum_{\r\in \mathcal{B}} \bhat_{f,\r}^{\dag}\bhat_{f,\r}\right]+\frac{U L}{4}
\label{eq:HcSPAM}
\end{align}
describes the effective charge dynamics of the SPAM, and
\begin{align}
\Hhat_{s} &= \frac{t\zeta_1}{4}\sum_{\r,\bmdelta} \bmsigma_\r\cdot\bmsigma_{\r+\bmdelta} +\frac{V\zeta_2}{2}\sum_{\r} \bmsigma_{\r}\cdot\bmtau_{\r}
\label{eq:HsSPAM}
\end{align}
is the model of its effective spin dynamics.  Here, $L$ is the total number of lattice sites and $\sfz$ is the nearest-neighbour coordination number. Note that in the $\Hhat_s$, the $\r$ is summed over the entire lattice, unlike in the $\Hhat_c$, where it is summed over %any 
one of the two sublattices. The $\bmdelta$ in both cases is summed over all the $\sfz$ nearest-neighbours. The real-valued decoupling parameters $\rho_1,~\rho_0,~\zeta_1$ and $\zeta_2$ are given by the following expectation values. 
\begin{subequations} \label{eq:paraSPAM}
 \begin{eqnarray}
  \rho_1 &=& \frac{1}{\sfz L}\sum_{\r, \bmdelta}\langle \bmsigma_{\r}\cdot\bmsigma_{\r+\bmdelta}\rangle \\
  \rho_0 &=& \frac{1}{L}\sum_\r \langle \bmsigma_{\r}\cdot\bmtau_{\r}\rangle \\
  \zeta_1 &=& \frac{2i}{\sfz L}\sum_{\r\in\mathcal{A}} \sum_\bmdelta\langle \phihat_{a,\r}\psihat_{b,\r+\bmdelta}\rangle \label{eq:zeta1SPAM} \\
  \zeta_2 &=& \frac{i}{L} \big\langle \sum_{\r\in\mathcal{A}}\phihat_{a,\r}\psitil_{a,\r} + \sum_{\r\in\mathcal{B}}\phitil_{b,\r}\psihat_{b,\r}\big\rangle \label{eq:zeta2SPAM}
 \end{eqnarray}
\end{subequations}
We compute these parameters self-consistently in the ground states of Eqs.~\eqref{eq:HcSPAM} and~\eqref{eq:HsSPAM}. In general, Eqs.~\eqref{eq:paraSPAM} are applicable at finite temperatures, but in this paper we study the zero temperature (ground state) properties only. 
We find the ground state of $\Hhat_c$ by numerical Bogoliubov diagonalization, and that of $\Hhat_s$ by doing triplon analysis in the bond-operator representation~\cite{SachdevBhatt1990,BK2010}. The details of these calculations are given below. Note that the effective spin dynamics of the SPAM is similar to that of the KLM~\cite{Ram2017}, except now the effective Kondo interaction, $V\zeta_2$, in Eq.~\eqref{eq:HsSPAM} is determined self-consistently. However, the charge dynamics of the SPAM is more complex compared to that of the KLM, because it involves the charge fluctuations of both $c$ as well as $f$ electrons.

%%%%%%%%%%%%%
\subsection{Effective charge dynamics \label{sec:charge-dynamics}}
To diagonalize the $\Hhat_c$ of Eq.~\eqref{eq:HcSPAM}, we first rewrite it in terms of the spinless fermion creation and annihilation operators using the definition of the Majorana operators given below Eq.~\eqref{eq:cf-kumar}. It reads as follows:  
\begin{align}
\Hhat_{c} =& -\frac{t}{2} \sum_{\r\in \mathcal{A}} \sum_\bmdelta \left[\rho_{1+} \,\ahat_{c,\r}^{\dag}\bhat^{ }_{c,\r+\bmdelta} + \rho_{1-} \ahat_{c,\r}^{\dag}\bhat_{c,\r+\bmdelta}^{\dag}+{\rm h.c.}\right] 
\nonumber \\
& -\frac{V}{2} \sum_{\r\in \mathcal{A}} \left[\rho_{0+} \, \ahat_{c,\r}^{\dag}\ahat^{ }_{f,\r} + \rho_{0-} \, \ahat_{c,\r}^{\dag}\ahat_{f,\r}^{\dag}+ {\rm h.c.} \right]  \nonumber \\
&-\frac{V}{2} \sum_{\r\in \mathcal{B}} \left[ \rho_{0+} \, \bhat_{f,\r}^{\dag}\bhat^{ }_{c,\r} + \rho_{0-} \bhat_{f,\r}^{\dag}\bhat_{c,\r}^{\dag}+ {\rm h.c.}\right] \nonumber \\
& - \frac{U}{2}\left[\sum_{\r\in \mathcal{A}} \ahat_{f,\r}^{\dag}\ahat^{ }_{f,\r} +\sum_{\r\in \mathcal{B}} \bhat_{f,\r}^{\dag}\bhat^{ }_{f,\r}\right] + \frac{U L}{4}
\end{align}
where $\rho_{1\pm}=(1\pm\rho_1)$ and $\rho_{0\pm}=(1\pm\rho_0)$. Then, by applying the Fourier transformation, $\ahat_{\vartheta,\r} = \sqrt{\frac{2}{L}}\sum_\k e^{i\k\cdot\r} \ahat_{\vartheta,\k} $ and $ \bhat_{\vartheta,\r} = \sqrt{\frac{2}{L}}\sum_\k e^{i\k\cdot\r} \bhat_{\vartheta,\k}$ (where $\vartheta=c,f$), we get $\Hhat_c = \sum_\k \Psi_\k^{\dag} \mathcal{H}^{ }_\k \Psi^{ }_\k $ in the $\k$-space, where the Nambu row-vector operator, $\Psi^\dag_\k$, is defined as:
\begin{align}
\Psi_\k^{\dag} & =\left[\ahat_{c,\k}^{\dag}~ \bhat_{c,\k}^{\dag}~ \bhat_{f,\k}^{\dag}~ \ahat_{f,\k}^{\dag}~ \ahat_{c,-\k}^{}~ \bhat_{c,-\k}^{}~ \bhat_{f,-\k}^{}~ \ahat_{f,-\k}^{}\right] 
\end{align}
and the corresponding column-vector, $ \Psi_\k = [\Psi_\k^\dag]^\dag$. Moreover, $\k\in$ half-BZ, $\gamma_\k=\sum_{\bmdelta}e^{i\k\cdot\r}=|\gamma_\k|~e^{i\varphi_\k}$, and a gauge transformation, $\ahat_c^{\dag}(\k)\to e^{-i\varphi_\k}\ahat_c^{\dag}(\k)$ and $\ahat_f^{\dag}(\k)\to e^{-i\varphi_\k}\ahat_f^{\dag}(\k)$, has been applied to absorb the phase $\varphi_\k$. The $\mathcal{H}_\k$ is the following $8\times8$ matrix:
\begin{equation}
\mathcal{H}_\k = \begin{bmatrix} 
 ~~A & ~~B \\
 -B & -A 
  \end{bmatrix}
  \label{eq:hkSPAM}
\end{equation}
with
\begin{subequations}
\begin{align}
& A =-\frac{1}{4}\begin{bmatrix} 
  0 & t|\gamma_\k|\rho_{1+} & 0 & V\rho_{0+}\\ 
  t|\gamma_\k|\rho_{1+} & 0 & V\rho_{0+} & 0\\ 
  0 & V\rho_{0+} & U & 0\\
  V\rho_{0+} & 0 & 0 & U
  \end{bmatrix}~\mbox{and} \\
& B =-\frac{1}{4}\begin{bmatrix} 
  0 & t|\gamma_\k|\rho_{1-} & 0 & V\rho_{0-}\\ 
  -t|\gamma_\k|\rho_{1-} & 0 & -V\rho_{0-} & 0\\ 
  0 & V\rho_{0-} & 0 & 0\\
  -V\rho_{0-} & 0 & 0 & 0
  \end{bmatrix}.
\end{align}
\end{subequations}

We diagonalize the $\Hhat_c$ by applying the Bogoliubov transformation on $\Psi_\k$. To do this, we define a unitary matrix, $\mathcal{U}_{\k}$, such that
\begin{align}
\mathcal{U}_{\k}^{\dag}\mathcal{H}_\k \mathcal{U}_{\k} & = \frac{1}{2}
\begin{bmatrix} 
\mathcal{E}_\k & 0\\
0 & -\mathcal{E}_\k
\end{bmatrix}
 \end{align}
where $\mathcal{E}_\k$ is a diagonal matrix with $E_{\k,1}$, $E_{\k,2}$, $E_{\k,3}$, $E_{\k,4}$ as its diagonal elements, and 
\begin{align}
& \Psi_\k^\dag \mathcal{U}_\k = \Lambda_\k^\dag \\
 & = \left[ \Lambda^\dag_{\k,1}~ \Lambda^\dag_{\k,2}~ \Lambda^\dag_{\k,3}~ \Lambda^\dag_{\k,4}~\Lambda^{ }_{-\k,1}~ \Lambda^{ }_{-\k,2}~ \Lambda^{ }_{-\k,3}~\Lambda^{ }_{-\k,4} \right] \nonumber 
\end{align} 
are the new canonical fermions describing the charge quasiparticles. In  terms of these quasiparticle operators, the $\Hhat_c$ is diagonal, and it reads as:
\begin{eqnarray}
\Hhat_c = \sum_\k\sum_{i=1}^{4}E_{\k,i}~ \biggl(\Lambda_{\k,i}^{\dag}\Lambda_{\k,i}^{}-\frac{1}{2}\biggr),
\label{eq:Hc-diag}
\end{eqnarray}
where $E_{\k,i}>0~,i=1,2,3,4$ are the dispersions of the charge quasiparticles.
The ground state of the $\Hhat_c$ is the vacuum state, $|\mathbf{0}_c\rangle$, of the $\Lambda_{\k,i}$ quasiparticles with ground state energy per site, $ e_{g,c} = - \frac{1}{2L} \sum_{\k,i} E_{\k,i}$.

In order to find the mean-field parameters $\zeta_1$ and $\zeta_2$, we rewrite Eqs.~\eqref{eq:zeta1SPAM} and~\eqref{eq:zeta2SPAM} in the $\k$-space, apply the Bogoliubov transformation, $\mathcal{U}_{\k}$, and then calculate the expectation values in the ground state of the $\Hhat_c$. Since $\langle \mathbf{0}_c|\Lambda_{\k,i}^{\dag}\Lambda_{\k,j}^{}|\mathbf{0}_c\rangle=0$ and $\langle \mathbf{0}_c|\Lambda_{\k,i}^{}\Lambda_{\k,j}^{\dag}|\mathbf{0}_c\rangle=\delta_{i,j}$, we get the following equations to compute $\zeta_1$ and $\zeta_2$.
\begin{subequations}
\label{eq:zeta1-zeta2}
\begin{eqnarray}
\zeta_1 &=&\frac{1}{\sfz L} \sum_\k\sum_{i=1}^{4} [\tilde{M}_{\zeta_1}(\k)]_{i+4,i+4} \\
\zeta_2 &=&\frac{1}{2L} \sum_\k\sum_{i=1}^{4} [\tilde{M}_{\zeta_2}(\k)]_{i+4,i+4}
\end{eqnarray}
\end{subequations}
Here, $ \tilde{M}_{\zeta_{1(2)}}(\k)= \mathcal{U}_{\k}^{\dag} M_{\zeta_{1(2)}}(\k)\mathcal{U}_{\k}$ with $M_{\zeta_1}(\k)$ and $M_{\zeta_2}(\k)$ as the following $8\times 8$ matrices.
\begin{align}
M_{\zeta_1}(\k)=&
\begin{bmatrix} 
 0&-|\gamma_\k|&0&0&0&|\gamma_k|&0&0\\
-|\gamma_\k|&0&0&0&-|\gamma_k|&0&0&0\\
 0&0&0&0&0&0&0&0\\
 0&0&0&0&0&0&0&0\\
 0&-|\gamma_\k|&0&0&0&|\gamma_k|&0&0\\
 |\gamma_\k|&0&0&0&|\gamma_k|&0&0&0\\
  0&0&0&0&0&0&0&0\\
 0&0&0&0&0&0&0&0
\end{bmatrix} \\
M_{\zeta_2}(\k)=& 
\begin{bmatrix} 
 0&0&0&-1&0&0&0&1\\
 0&0&-1&0&0&0&-1&0\\
 0&-1&0&0&0&1&0&0\\
 -1&0&0&0&-1&0&0&0\\
 0&0&0&-1&0&0&0&1\\
 0&0&1&0&0&0&1&0\\
 0&-1&0&0&0&1&0&0\\
 1&0&0&0&1&0&0&0
\end{bmatrix}
\end{align}
%%%%%%%%%%%%%%%
%%%%%%%%%%%%%%%%
\subsection{Effective spin dynamics \label{sec:spin-dynamics}}
We study the spin dynamics of the SPAM, given by the $\Hhat_{s}$ of Eq.~\eqref{eq:HsSPAM}, by doing the bond-operator mean-field theory, as we did in Ref.~\onlinecite{Ram2017} for the KLM. Since every lattice site here has a pair of spin-1/2 operators $\bmsigma_\r$ and $\bmtau_\r$, we can use the bosonic bond-operators, $\shat_\r$ and $\that_{\r\alpha}$, corresponding respectively to the local singlet and triplet states with a physical constraint: $\shat^{\dagger}_\r\shat^{ }_\r+\sum_\alpha\that^{\dagger}_{\r\alpha}\that^{ }_{\r\alpha}=1$,  to describe the two %sets of 
spin-1/2 operators as~\cite{SachdevBhatt1990}:
\begin{subequations}
\label{eq:bop}
\begin{align}
\bmsigma^\alpha_{\r} & = \left(\shat_\r^\dag\that^{ }_{\r\alpha} + {\rm h.c.}\right) -i\epsilon_{\alpha\beta\gamma}\that^\dag_{\r\beta}\that^{ }_{\r\gamma} \\
\bmtau^\alpha_{\r} & = - \left(\shat_\r^\dag\that^{ }_{\r\alpha} + {\rm h.c.}\right) -i\epsilon_{\alpha\beta\gamma}\that^\dag_{\r\beta}\that^{ }_{\r\gamma}
\end{align}
\end{subequations}
where $\alpha = x,y,z$ denote their three components (and likewise, $\beta$ and $\gamma$) and $\epsilon_{\alpha\beta\gamma}$ is the Levi-Cevita tensor. 

Since the effective Kondo interaction, $V\zeta_2$, facilitates the formation of local singlet between $\bmsigma_\r$ and $\bmtau_\r$, we formulate a bond-operator mean-field theory of the $\Hhat_s$ with respect to this singlet state. In this theory, the reference Kondo singlet state is described by the mean singlet amplitude: $\langle \shat_\r \rangle \approx \sbar$, while the triplet fluctuations on top of it are treated quantum mechanically. For further simplification, the interaction between the triplet excitations is also neglected. These approximations basically amount to rewriting Eqs.~\eqref{eq:bop} as: $\bmsigma^{\alpha}_\r \approx \sbar \left(\that^{ }_{\r\alpha}+\that^{\dagger}_{\r\alpha}\right) \approx -\bmtau^{\alpha}_\r $. Moreover, the exchange interaction between $\bmsigma_\r$ and $\bmtau_\r$ reads as: 
$ \bmsigma_\r \cdot \bmtau_\r \approx-3\sbar^2 + \sum_\alpha \that^{\dagger}_{\r\alpha}\that^{ }_{\r\alpha}$.

Under these approximations, the $\Hhat_s$ takes the following mean-field form:
\begin{align}
\Hhat_s=&~ \frac{t\zeta_1\sbar^2}{4} \sum_{\r,\bmdelta,\alpha} \left(\that^{\dagger}_{\r,\alpha}+\that_{\r,\alpha}\right) \left(\that^{\dagger}_{\r+\bmdelta,\alpha}+\that_{\r+\bmdelta,\alpha}\right) \nonumber \\
& + \frac{V\zeta_2}{2}\sum_\r \Big(-3\sbar^2+ \sum_\alpha \that^{\dagger}_{\r,\alpha}\that_{\r,\alpha} \Big) \nonumber \\ 
&-\lambda\sum_\r \Big(\sbar^2+\sum_\alpha \that^{\dagger}_{\r,\alpha}\that_{\r,\alpha}-1\Big), 
\end{align}
where the Lagrange multiplier, $\lambda$, is introduced to satisfy the constraint on average. %Under 
After doing the Fourier transformation, $\that_{\r\alpha} = \frac{1}{\sqrt{L}}\sum_{\k}e^{i\k\cdot\r}\that_{\k\alpha}$, it reads as:
\begin{align} \label{eq:hmf_kspc}
\Hhat_s = & \frac{1}{2}\sum_{\k,\alpha}\Bigg\{\left[\lambda+\frac{1}{2}t\zeta_1\sbar^2\gamma_{\k} \right] \left(\hat t^\dag_{\k\alpha} \hat t_{\k \alpha} + \hat t_{-\k\alpha}\hat t^\dag_{-\k\alpha}\right)
\nonumber \\
& +\frac{1}{2}t\zeta_1\sbar^2\gamma_{\k} \left(\hat t^\dag_{\k\alpha}\hat t^\dag_{-\k\alpha} + \hat t_{-\k\alpha}\hat t_{\k\alpha}\right) \Bigg\} +  e_0 L
\end{align}
where $\k\in$ the full BZ, $\lambda\to (V\zeta_2/2-\lambda)$ is the effective chemical potential of triplons, $\gamma_\k=\sum_\bmdelta e^{i\k\cdot\bmdelta}$, and $e_0=\left[\lambda \sbar^2-\frac{5}{2}\lambda -2V\zeta_2 \left(\sbar^2-1/4\right)\right]$. This triplon Hamiltonian can be diagonalized by the Bogoliubov transformation:
\begin{eqnarray}
 \that^{ }_{\k\alpha}&=&\ttilde^{ }_{\k\alpha}\cosh {\theta_{\k}} - \ttilde_{-\k\alpha}^{\dagger}\sinh \theta_{\k}
  \label{eq:Blg_trsfm}
\end{eqnarray}
for $\theta_\k = \frac{1}{2} \tanh^{-1}\left[\frac{\frac{1}{2}t\zeta_1\sbar^2\gamma_{\k}}{(\lambda+\frac{1}{2}t\zeta_1\sbar^2\gamma_{\k})}\right]$. The $\ttilde_{\k\alpha}$'s are the new bosonic operators describing the triplon quasiparticles with dispersion, $\varepsilon_\k = \sqrt{\lambda(\lambda+t\zeta_1\sbar^2 \gamma_\k)} \ge 0$, in term of which the diagonalized $\Hhat_s$ reads as follows.
\begin{eqnarray}
\Hhat_s& =& e_0 L + \sum_{\k,\alpha} \varepsilon_{\k}\left(\ttilde^\dag_{\k\alpha}\ttilde^{ }_{\k\alpha} +\frac{1}{2}\right)
\label{eq:Hs-diag}
\end{eqnarray}
Its ground state energy per site is given as: $ e_{g,s}[\lambda,\sbar^2]= e_0 +\frac{3}{2L}\sum_{\k}\varepsilon_{\k} $. By minimizing $e_{g,s}$ with respect to $\lambda$ and $\sbar^2$, we get the following equations whose solution determines $\lambda$ and $\sbar^2$, and from which the decoupling parameters for the spin part can be obtained as: $\rho_0 = 1-4\sbar^2$ and $\rho_1 = 4\sbar^2(2V\zeta_2 - \lambda)/\sfz t\zeta_1$.\begin{subequations}
\label{eq:sbar-lambda}
\begin{eqnarray}
\sbar^2 &=& \frac{5}{2}-\frac{3}{4L}\sum_\k\frac{2\lambda+t\zeta_1\sbar^2 \gamma_\k}{\varepsilon_\k} \\
\lambda &=& 2V\zeta_2 - \frac{3\lambda t \zeta_1}{4L}\sum_\k \frac{\gamma_\k}{\varepsilon_\k}
\end{eqnarray}
\end{subequations}

We determine $\zeta_1,~\zeta_2,~\rho_0$ and $\rho_1$ defined in Eqs.~\eqref{eq:paraSPAM} by numerically solving the Eqs.~(\ref{eq:zeta1-zeta2}) and~(\ref{eq:sbar-lambda}). In the following sections, we discuss the physical behaviour of the SPAM as obtained from this self-consistent theory. 
\begin{figure}[!htb]
\centering
\includegraphics[width=.43\textwidth]{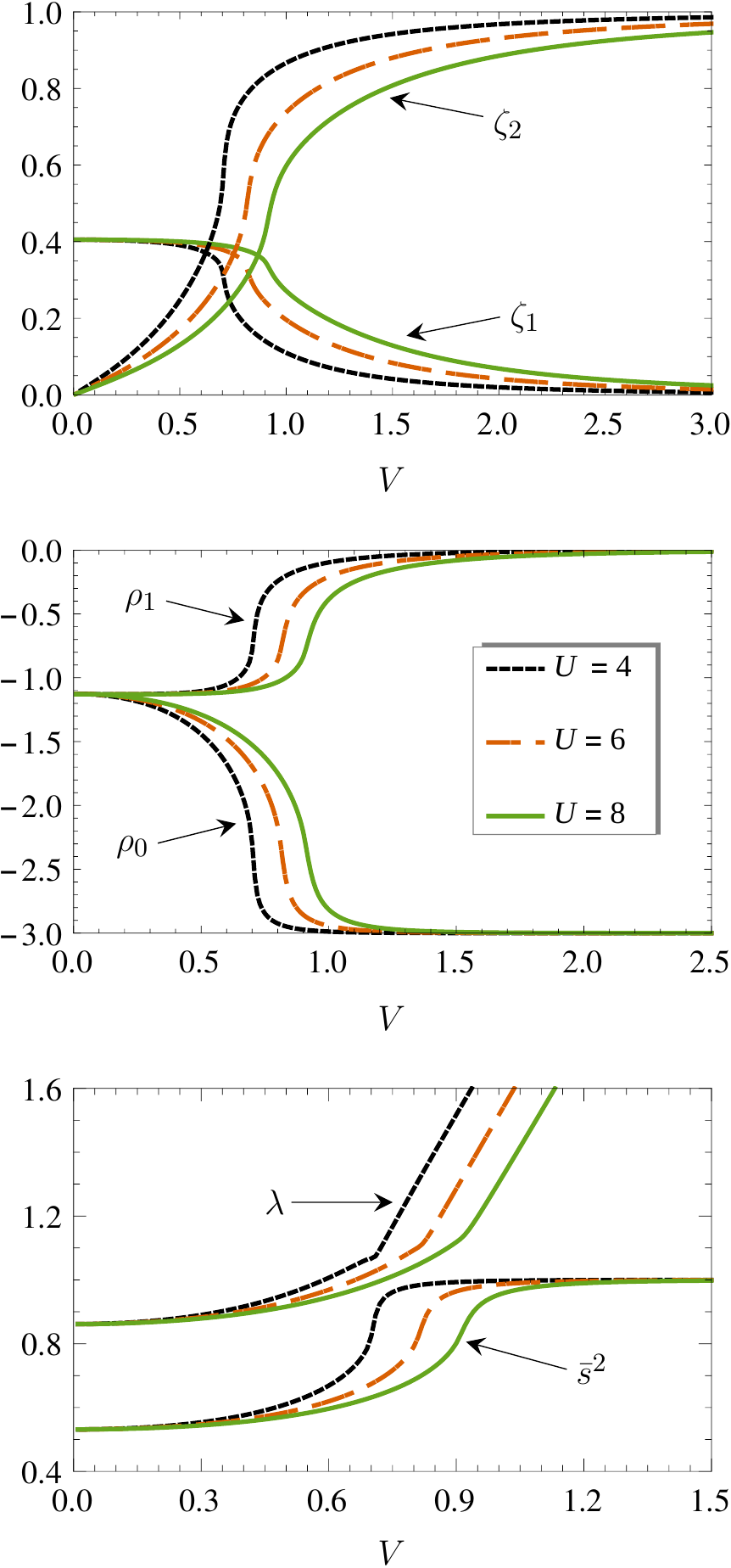}
\caption{The mean-field parameters, $\zeta_1,~\zeta_2,~\rho_0,~\rho_1,~\lambda$ and $\sbar^2$, of the charge and spin dynamics on the square lattice, as a function of $V$ for $U=4,6$ and $8$ (given by the same legend in all three plots). For large $V$, we get $\sbar^2 \rightarrow 1$ and $\rho_0 \rightarrow -3$, which implies a perfect Kondo singlet.} 
\label{fig:mean-field}
\end{figure}

%%%%%%%%%%%%%%%%%%%%%
\section{Magnetic Transition in the Ground State}\label{sec:ground-state-results}
We investigate the ground state properties of the SPAM by self-consistently solving the Eqs.~(\ref{eq:zeta1-zeta2}) and~(\ref{eq:sbar-lambda}) on the square lattice. In our calculations, we put $t=1$, and compute the parameters of the effective charge and spin dynamics as a function of $V$ for different values of $U$. The data thus obtained for various mean-field parameters 
is plotted in Fig.~\ref{fig:mean-field}. Notably, for large values of $V$, we get $\sbar^2\rightarrow 1$ and $\rho_0 \rightarrow -3$, which correctly implies a perfect Kondo singlet state. However, as $V$ decreases, $\sbar^2$ also decreases, and at a $U$ dependent critical point, $V_c$, the Kondo singlet phase undergoes a continuous transition to the N\'eel AFM phase. To this end, we calculate the triplon dispersion, and follow the spin gap in the $U$-$V$ plane to generate the quantum phase diagram. We also compute the charge quasiparticle dispersions, which show the ground state to be insulating, and display two inversion transitions. But first, we discuss the magnetic transition in the ground state. 

%%%%%%%%%%%%%%%%%%%
\subsection{Triplon dispersion and spin gap}
\begin{figure}[b]
\centering
\includegraphics[width=.45\textwidth]{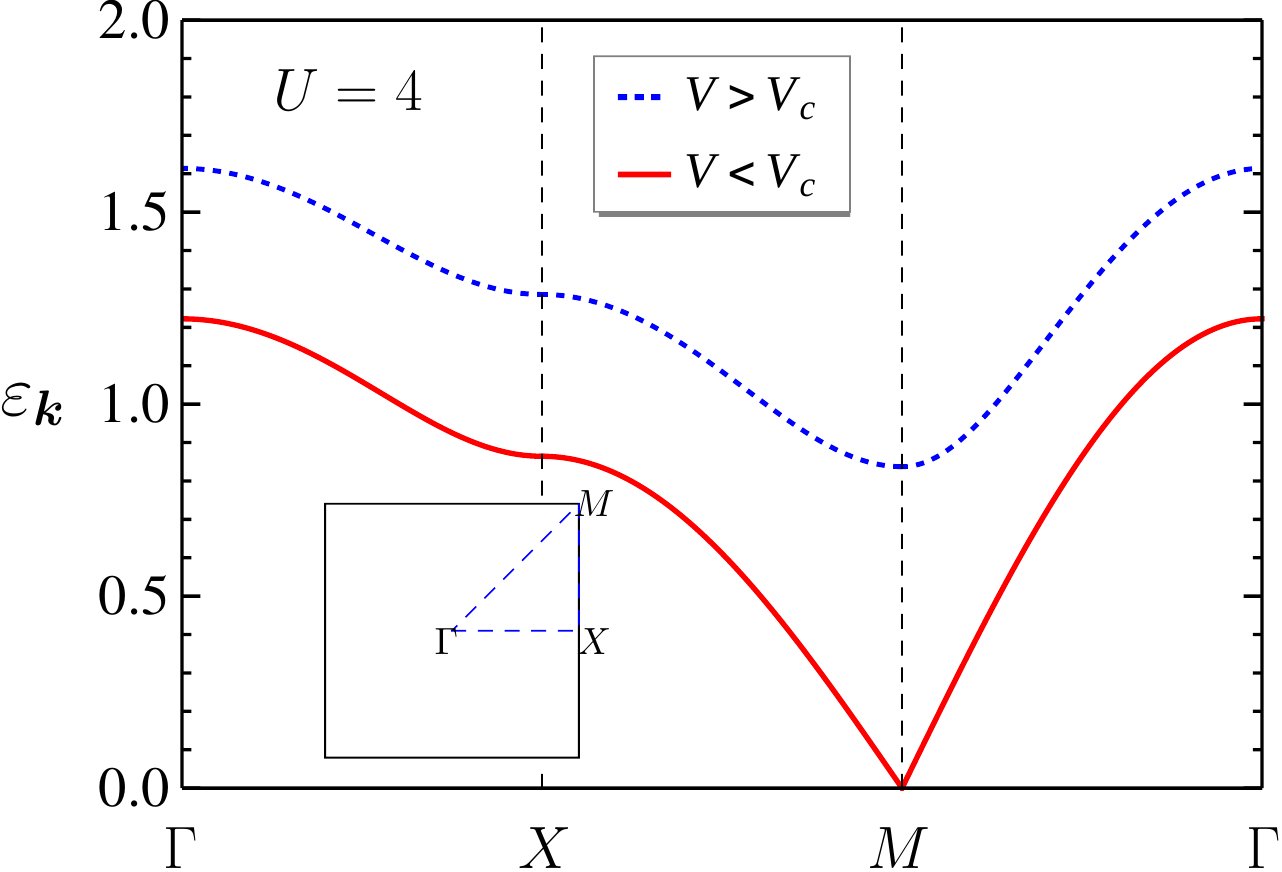}
\caption{Triplon dispersion, $\varepsilon_\k$, in the Brillouin zone of square lattice, in Kondo singlet (blue dashed curve) and AFM phases (red solid curve) at $U=4$. It is gapped for $V>V_c$, and gapless for $V<V_c$, where $V_c$ marks the onset of gaplessness.}
\label{fig:spin-disp}
\end{figure}

We compute the triplon dispersion, $\varepsilon_\k$, as given in Eq.~(\ref{eq:Hs-diag}). It is found to have an energy gap for large values of $V$ for any $U$. It remains gapped with decreasing $V$ upto a critical value, $V_c$. Thus, for $V>V_c$, the system is in the spin-gapped Kondo singlet phase. At $V_c$, however, the $\varepsilon_\k$ becomes gapless at $\k=\Q=(\pi,\pi)$, and stays gapless for $V < V_c$. The gapless nature of $\varepsilon_\k$ at $\Q$ implies  Bose condensation of triplons, which in turn implies N\'eel antiferromagnetic order. This phase transition by decreasing $V$ occurs for any $U$, but at a $V_c$ which depends upon $U$. In Fig.~\ref{fig:spin-disp}, we have plotted the the triplon dispersion obtained from our self-consistent calculations for two different values of $V$ on both sides of $V_c$ for $U=4$.
\begin{figure}[!htb]
\centering
\includegraphics[width=.45\textwidth]{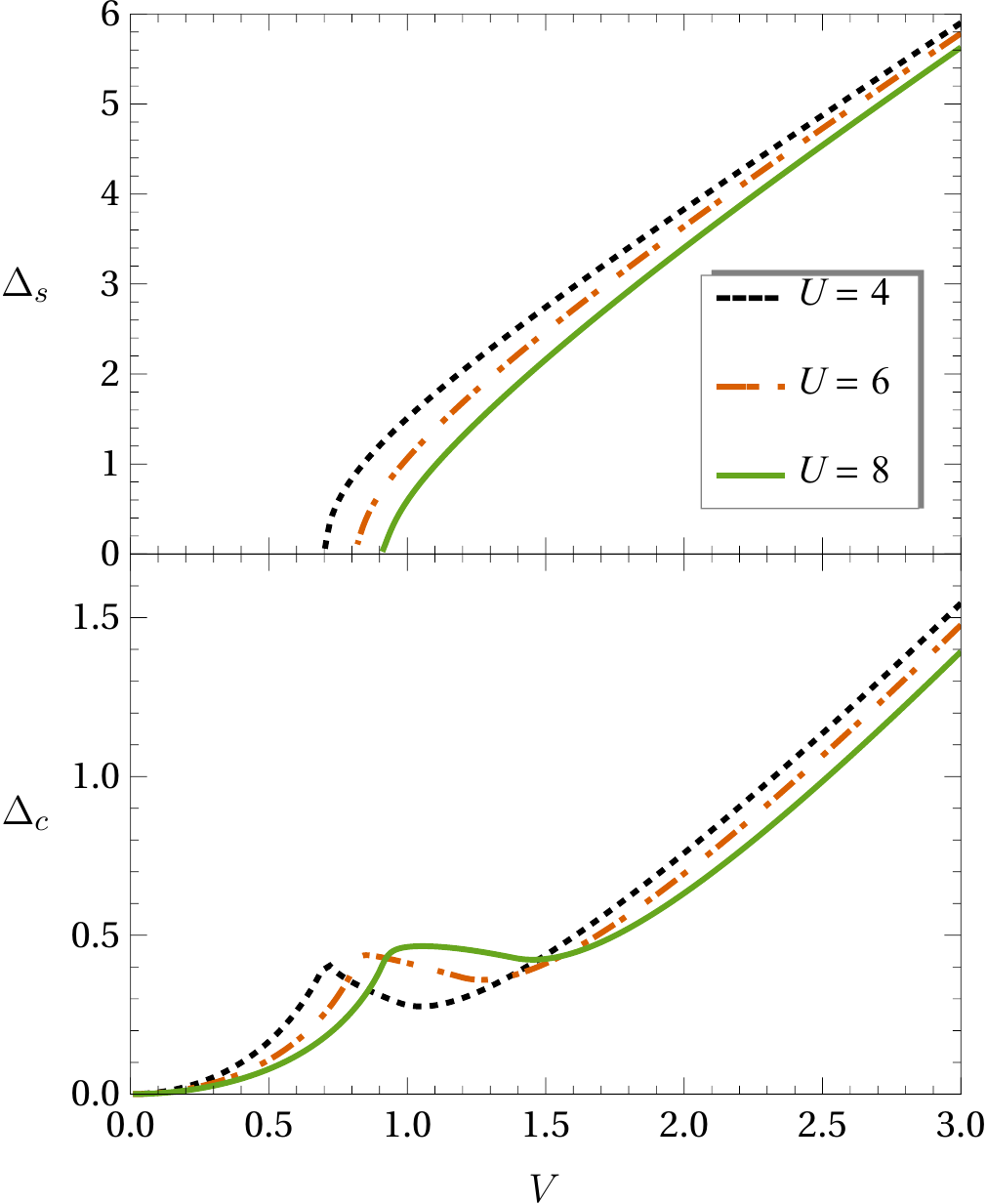}
\caption{The spin gap, $\Delta_s$, and the charge gap, $\Delta_c$, vs. $V$ for $U=4,6$ and $8$, on square lattice. The $\Delta_s$ vanishes continuously as $V$ approaches the critical point, $V_c$, implying a continuous transition from the gapped Kondo singlet to the gapless AFM phase in the ground state. The charge gap, $\Delta_c$, is always non-zero implying an insulating ground state.}
\label{fig:spin-gap}
\end{figure}

From the triplon dispersion, we calculate the spin gap, $\Delta_s$, as a function of $V$ for different values of $U$. Since the minimum of $\varepsilon_\k$ is always at $\Q$, the spin gap is given by the equation $\varepsilon_\Q=\Delta_s$. That is, $\Delta_s=\sqrt{\lambda(\lambda-\sfz t\zeta_1\sbar^2)}$. The calculated spin gap, shown in Fig.~\ref{fig:spin-gap} (top panel), vanishes continuously at $V_c$, which implies a continuous phase transition in the ground state. We observe that as $U$ increases, $V_c$ also increases. Hence, the $U$ supports the AFM order, while the $V$ favours the Kondo singlet. For comparison, in the bottom panel of Fig.~\ref{fig:spin-gap}, we also present the charge gap, $\Delta_c$ (from our calculations discussed in the next section). Notably, the $\Delta_c$ is always non-zero implying an insulating state.
 
%%%%%%%%%%%%%%%%%%
\subsection{Quantum phase diagram}
We obtain the phase boundary between the Kondo singlet and the N\'eel AFM phases in the $U$-$V$ plane by meeting the condition, $\varepsilon_\Q=0$, from the gapped side. It marks the instability of the Kondo singlet phase towards magnetic ordering. It fixes the $\lambda$ of the bond-operator mean-field theory as: $\lambda = \sfz t\zeta_1 \sbar^2$. With this value of $\lambda$, after a few steps of manipulation of Eqs.~\eqref{eq:sbar-lambda} that apply to the gapped (Kondo singlet) phase, we get the following equation for the critical hybridization.
\begin{eqnarray}
V_c &=& \frac{1}{2}\sfz t(\sbar^2+3y)\left(\frac{\zeta_1}{\zeta_2}\right)
\label{eq:Vc}
\end{eqnarray}
The self-consistent parameters of the spin part are found to become constants at the phase boundary. They are $\rho_0= 1-4\sbar^2$, $\rho_1= 12\sbar^2 y$ and $\sbar^2=5/2 - 3x$, where
$$x=\frac{1}{4L}\sum_\k \frac{2+(\gamma_\k/\sfz)}{\sqrt{1+(\gamma_\k/\sfz)}}~\text{and}~y=\frac{1}{4L}\sum_\k \frac{(\gamma_\k/\sfz)}{\sqrt{1+(\gamma_\k/\sfz)}}$$ are two constants. 
\begin{figure}[!htb]
\centering
\includegraphics[width=.45\textwidth]{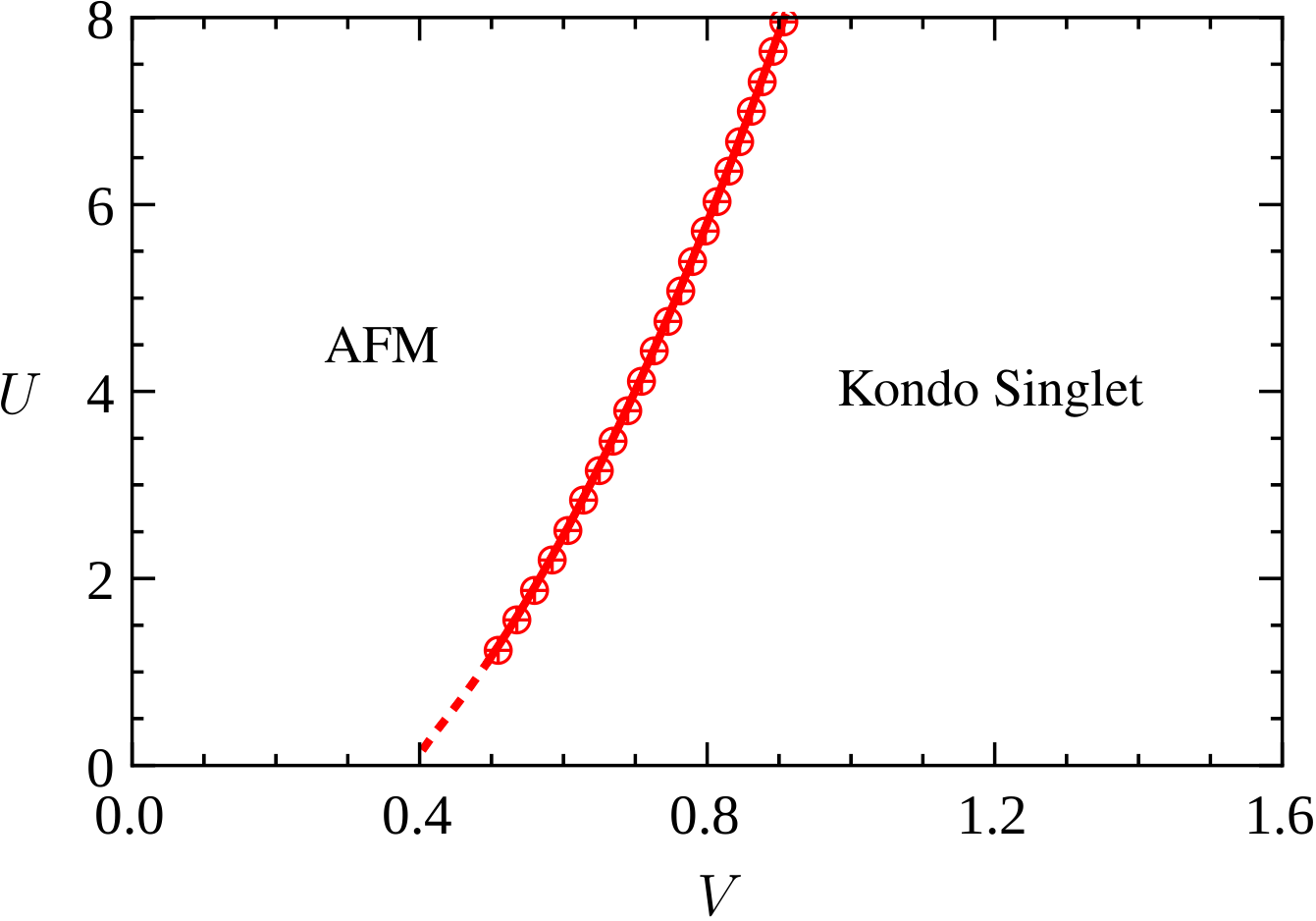}
\caption{Quantum phase diagram of the symmetric periodic Anderson model on the square lattice in the $U$-$V$ plane (with $t=1$), as obtained from our theory. From the large to moderate values of $U$, it agrees qualitatively with the QMC  phase boundary~\cite{Vekic1995}. But for small $U$'s, it does not produce a $V_c$ (dashed line) that is expected to tend to $0$ from the mean-field treatment of the local repulsion.}
\label{fig:phasediag2}
\end{figure}
So, the critical hybridization, $V_c$, depends implicitly upon $U$ through the parameters $\zeta_1$ and $\zeta_2$ of the charge part.

We calculate $V_c$ as a function of $U$ by numerically solving Eq.~\eqref{eq:Vc} together with Eqs.~(\ref{eq:zeta1-zeta2}) and~(\ref{eq:sbar-lambda}). It gives us the phase boundary between the quantum paramagnetic Kondo singlet phase and the N\'eel ordered antiferromagnetic phase in the $U$-$V$ plane. The resulting quantum phase diagram is shown in Fig.~\ref{fig:phasediag2}. We see that the $V_c$ increases with $U$. This is consistent with the fact that the effective Kondo exchange interaction in the SPAM is $J\sim V^2/U$~\cite{Schrieffer1966,Sun1995}. So, an increase in $U$ reduces the strength of the effective Kondo coupling that allows the AFM order to survive upto the correspondingly larger value of $V_c$. From the moderate to large values of $U$, our theory produces a phase boundary that is in qualitative agreement with quantum monte carlo (QMC) calculations~\cite{Vekic1995}. However, for small $U$'s, the mean-field theory with N\'eel order is known to give a $V_c$ that rapidly vanishes as $U$ goes to zero~\cite{Moller1993}, whereas the $V_c$ from our calculations tends to a non-zero value even when $U$ becomes zero. It may be emphasized here that our theory, by construction, is better suited for larger values of $U$.
\begin{figure*}[!htb]
\centering
\includegraphics[width=\textwidth]{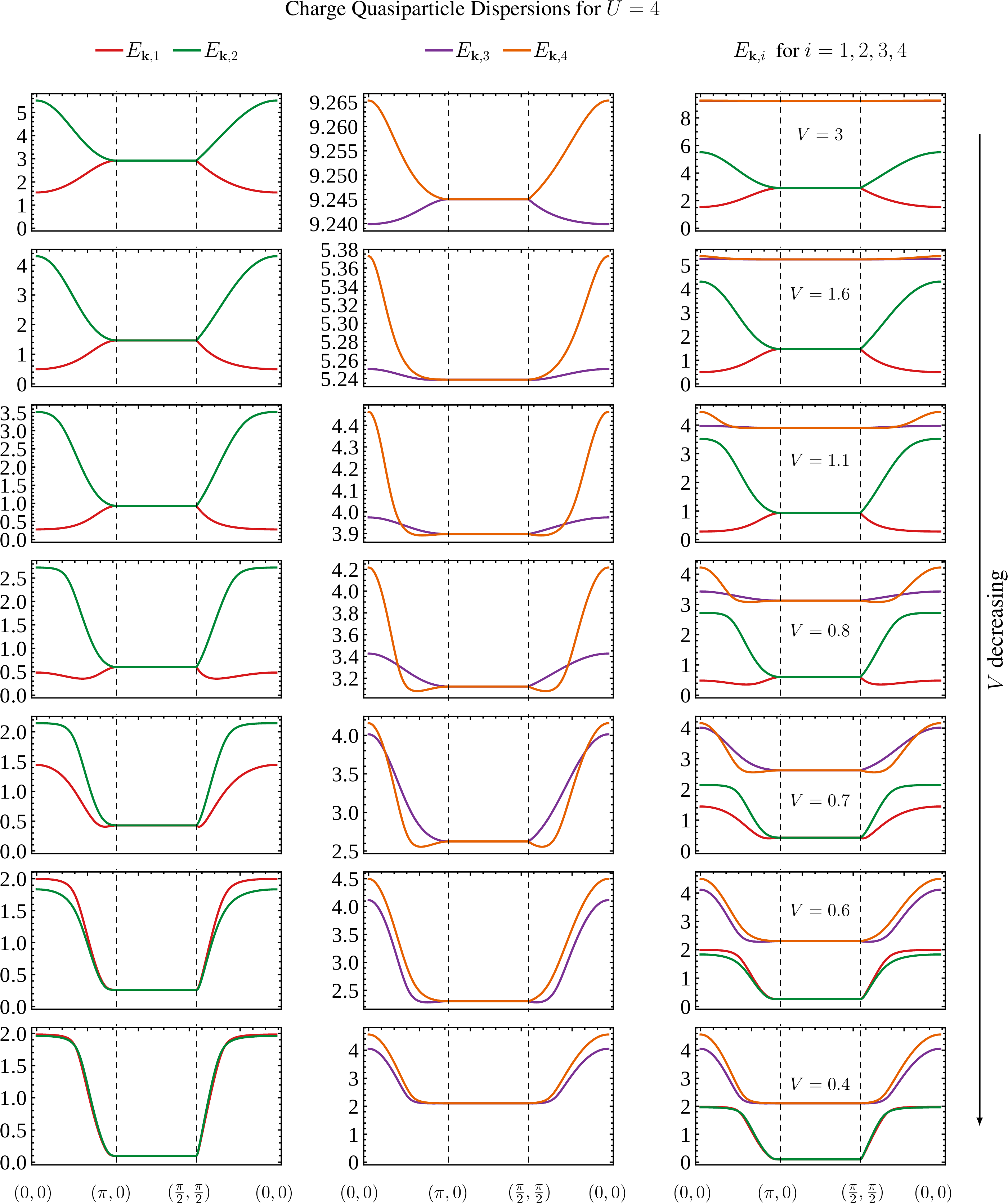} 
\caption{The evolution of the charge quasiparticles' dispersions with respect to $V$ on the square lattice. The dispersions, $\{E_{\k,1}$, $E_{\k,2}$, $E_{\k,3}$, $E_{\k,4}\}$, are plotted along the high symmetry lines in the half Brillouin zone. We see the inversion happening first for $E_{\k,3}$ (second column; below $V_{i3}=2.41$) and then for $E_{\k,1}$ (first column; below $V_{i1} = 0.96$) as $V$ decreases. In the third column, all the four bands are plotted together. Note that the higher energy narrow bands become broader upon decreasing $V$.}
\label{fig:charge-disp}
\end{figure*}

%%%%%%%%
\section{Inversion Transitions for the Charge Quasiparticles \label{sec:2inversions}}
Now let us discuss the nature of the charge quasiparticles in our theory of the SPAM. To this end, we compute the dispersions, $E_{\k,i}$ (for $i=1,2,3,4$), of the charge quasiparticles, as defined in Eq.~(\ref{eq:Hc-diag}). These $E_{\k,i}$'s are the positive eigenvalues of the matrix $\mathcal{H}_\k$ of Eq.~\eqref{eq:hkSPAM}. The evolution of the charge quasiparticles' dispersions with $V$ on the square lattice is presented in Fig.~\ref{fig:charge-disp}. Note that, for any non-zero $V$, the pair of quasiparticle bands $E_{\k,1}$ and $E_{\k,2}$ is always lower by a finite energy than the pair $E_{\k,3}$ and $E_{\k,4}$. Since the lowest dispersion, $E_{\k,1}$, is strictly non-zero for any $V\neq0$, it implies an insulating ground state with a finite charge gap (see the $\Delta_c$ in Fig.~\ref{fig:spin-gap}). 

For large values of $V$, we find the band $E_{\k,1}$ ($E_{\k,2}$) to have the minimum (maximum) value at $\k=(0,0)$ (the $\Gamma$ point), and the maxima (minima) at the $|\gamma_\k|=0$ contour where the two bands touch each other [see the middle branch from $(\pi,0)$ to $(\pi/2,\pi/2)$ in Fig.~\ref{fig:charge-disp}, which lies on the boundary of the half BZ of the square lattice]. Thus, they are mutually oppositely oriented. The bands $E_{\k,3}$ and $E_{\k,4}$ look similar, except that they are very narrow compared to $E_{\k,1}$ and $E_{\k,2}$ for large values of $V$. However, upon decreasing $V$, they tend to become broader. Eventually, for sufficiently small values of $V$, all the four bands have comparable bandwidths. But something even more important happens upon decreasing $V$, and that is the inversion of two of these bands. 

\begin{figure}[t]
\centering
\includegraphics[width=.45\textwidth]{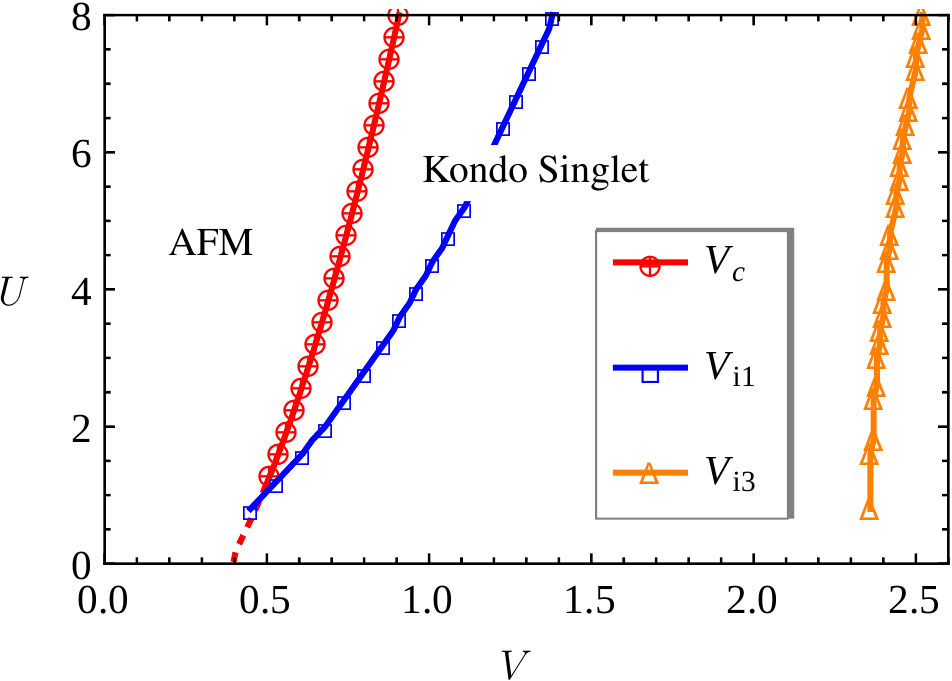}
\caption{The inversion transition lines $V_{i1}$ (blue squares) and $V_{i3}$ (orange triangles) correspond respectively to the charge quasiparticle bands $E_{\k,1}$ and $E_{\k,3}$ of the SPAM (see Fig.~\ref{fig:charge-disp}). For comparison, also shown here is the phase boundary (red circles) between the Kondo singlet and N\'eel AFM phases.}
\label{fig:inversion}
\end{figure}

Upon decreasing the hybridization, we see two inversion transitions occurring separately for a narrow and a broad charge quasiparticle band. In particular, we find that as soon as $V$ becomes smaller than a characteristic value, $V_{i3}$, the $\Gamma$ point becomes a point of local maxima for $E_{\k,3}$, and its minimum shifts onto a contour around the $\Gamma$ point. To see this, take a look at the purple coloured dispersion curves in the top two plots of
the second column in Fig.~\ref{fig:charge-disp}. While the $E_{\k,3}$ is undergoing inversion, the $E_{\k,4}$ shows no such change. Neither do $E_{\k,1}$ and $E_{\k,2}$ show any qualitative change across $V_{i3}$, but only for a while! As we reduce the hybridization further, there comes a second inversion point, $V_{i1}$, at which the lowest band, $E_{\k,1}$, starts inverting by shifting its minimum away from the $\Gamma$ point. Look carefully at the plots of the first column in Fig.~\ref{fig:charge-disp}. Finally, for the $V$'s sufficiently less than $V_{i1}$, both $E_{\k,1}$ and $E_{\k,3}$ are fully inverted and look pretty much like their respective partners, $E_{\k,2}$ and $E_{\k,4}$. Hence, as for the KLM in Ref.~\onlinecite{Ram2017}, the charge quasiparticle bands of the SPAM also undergo inversion transition. But here  the inversion is richer by two! That is, the SPAM exhibits two inversion transitions, first for a narrow high energy charge quasiparticle band, and then for the lowest energy charge quasiparticle band. This is a novel finding for the symmetric periodic Anderson model. In Fig.~\ref{fig:inversion}, we show the inversion transition lines corresponding to $V_{i1}$ and $V_{i3}$ as obtained from our calculations, together with the magnetic phase boundary, in the $U$-$V$ plane. 

We end this section with a remark on the charge gap, $\Delta_c$. The bottom panel of Fig.~\ref{fig:spin-gap} shows $\Delta_c$ as a function of $V$ for three different values of $U$. For strong hybridization, the charge gap comes from the $\Gamma$ point where $E_{\k,1}$ is minimum. That is, $\Delta_c = E_{(0,0),1}$. But for $V<V_{i,1}$, due to the inversion of $E_{\k,1}$, the $\Delta_c$ comes from a contour around the $\Gamma$ point. This contour, upon decreasing $V$, tends towards the boundary of the half Brillouin zone. The behaviour of the charge gap here is similar to what we had found for the KLM~\cite{Ram2017,Faye2018}.  

%%%%%%%%%
\section{Quantum Oscillations of Magnetization}\label{sec:QOs}
Our past experience with the KLM shows that, with inverted charge quasiparticle bands, the magnetic quantum oscillations show up nicely even in the insulating state. Since we do find the inversion to occur for the charge quasiparticles of the SPAM, we are hopeful of seeing the dHvA oscillations here too. Therefore, we investigate the quantum oscillations of magnetization in the ground state of the SPAM. For this purpose, we study the orbital response of Eq.~\eqref{eq:SPAM} with Peierls coupling to a uniform magnetic field via the hopping term which now carries a phase factor involving the vector potential, $\A$, and reads as:  $-t\sum_{\r,\bmdelta,s} e^{i\frac{e}{\hbar}\int_\r^{\r+\bmdelta}\A\cdot \dr} \, \chat^\dag_{\r,s} \chat^{ }_{\r+\bmdelta,s}$. We take $\A = -By\xhat$, for the magnetic field, $B$, along the $\zhat$ direction. By rewriting the SPAM with Peierls coupling in the representation of Eq.~\eqref{eq:cf-kumar}, we derive the following minimal effective model of the field dependent charge dynamics.
\begin{align}
\Hhat_{c}^{[B]} = &  -\frac{it}{2} \sum_{\r\in \mathcal{A}} \sum_\bmdelta \biggl\{  \cos(2\pi\alpha\r_y\hat{x}\cdot\hat{\bmdelta}) \Bigl[\psihat_{a,\r}\phihat_{b,\r+\bmdelta} + \nonumber  \\
& \rho_1 \psihat_{b,\r+\bmdelta}\phihat_{a,\r}\Bigr] \biggr\}  -\frac{iV}{2} \sum_{\r\in \mathcal{A}} \left[\psihat_{a,\r}\phitil_{a,\r}+\rho_{0}\psitil_{a,\r}\phihat_{a,\r}\right] \nonumber  \\
& -\frac{iV}{2} \sum_{\r\in \mathcal{B}} \left[\psitil_{b,\r}\phihat_{b,\r}+\rho_{0}\psihat_{b,\r}\phitil_{b,\r}\right]\nonumber\\
&-\frac{U}{2}\left[\sum_{\r\in \mathcal{A}} \ahat_{f,\r}^{\dag}\ahat_{f,\r} + 
\sum_{\r\in \mathcal{B}} \bhat_{f,\r}^{\dag}\bhat_{f,\r}\right]
\label{eq:HcSPAM-B}
\end{align}
This is similar to what we had obtained for the KLM~\cite{Ram2017}. Here, $\alpha=e B a^2/h$ is the reduced magnetic flux, with $a$ as the lattice constant. The $\r_y$ and $\hat{\bmdelta}=\bmdelta/|\bmdelta|$ are the $y$ coordinate of $\r$ and unit vector for $\bmdelta$, respectively. For $\alpha=0$, Eq.~\eqref{eq:HcSPAM-B} reduces to Eq.~\eqref{eq:HcSPAM}. 

\begin{figure}[htbp]
\centering
\includegraphics[width=.45\textwidth]{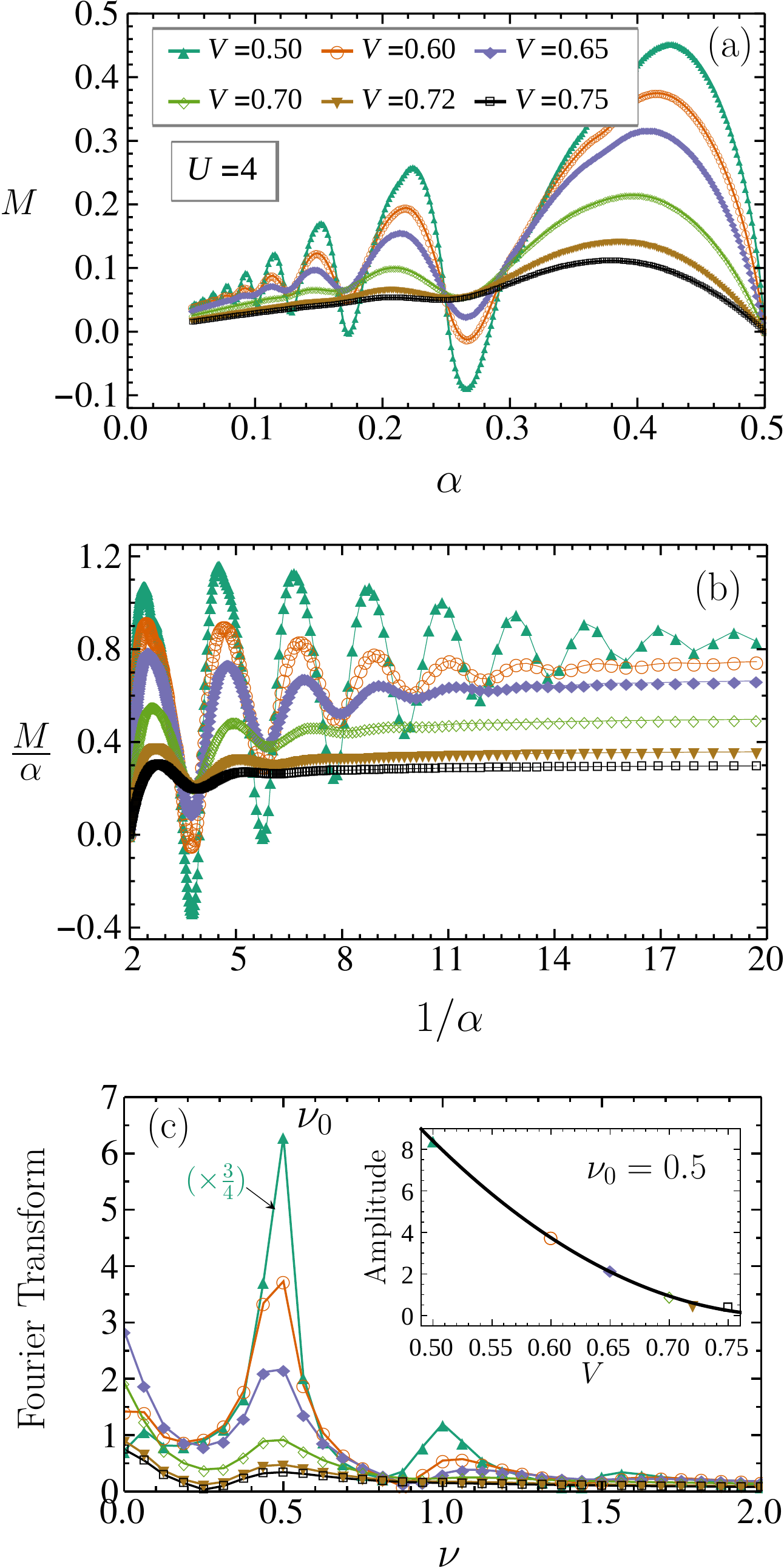}
\caption{The dHvA oscillations in the insulating ground state of the symmetric periodic Anderson model on square lattice, for $V=0.75,0.72,0.7$ (Kondo singlet) and $0.65,0.6,0.5$ (N\'eel AFM) for $U=4$. The legend in (a) is common to all the plots. The frequency, $\nu_0=0.5$, corresponds to the area of the half Brillouin zone. The inset of (c) shows the Fourier amplitude of the $\nu_0$ peak, with an empirical fit to $93.37(V-0.8)^2$.}
\label{fig:QOs1}
\end{figure}

\begin{figure}[htbp]
\centering
\includegraphics[width=.45\textwidth]{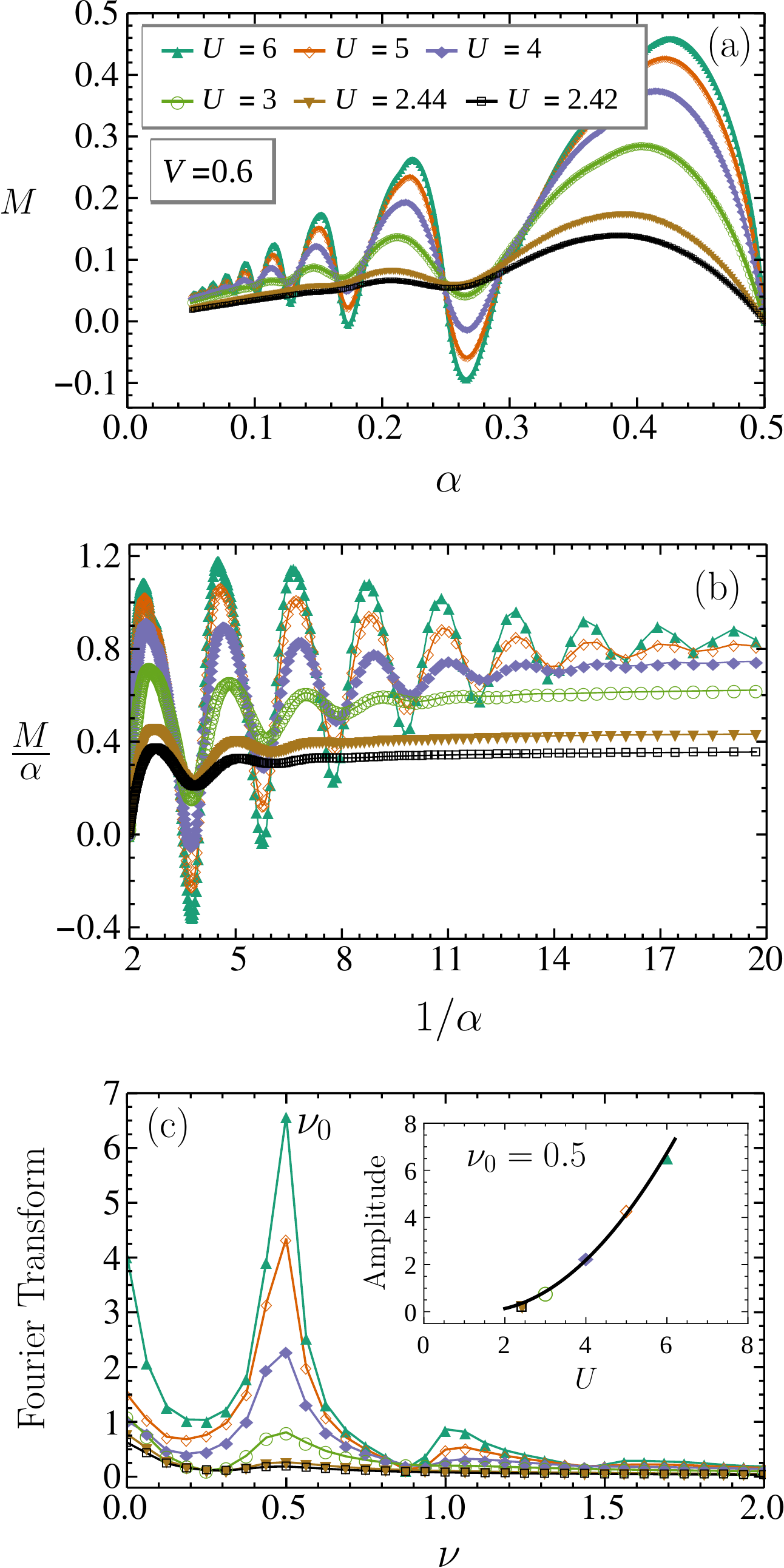}
\caption{The dHvA oscillations in the insulating ground state of the symmetric periodic Anderson model on square lattice for $U=2.42$, 2.44 (Kondo singlet) and $U=3,4,5,6$ (N\'eel AFM) for $V = 0.6$. The inset of (c) shows the Fourier amplitude of the $\nu_0$ peak, with an empirical fit to $0.31 (U-1.35)^2$.}
\label{fig:QOs2}
\end{figure}

To calculate the magnetization, $M$, versus $B$ from this Hofstadter like problem, we put the zero-field values of  $\rho_0$ and $\rho_1$ (see Fig.~\ref{fig:mean-field}) in Eq.~(\ref{eq:HcSPAM-B}), and numerically compute its ground state energy per site, $e_g$, as a function of $\alpha=p/q$ for integer $p=1,2,\dots,q$ on the square lattice. We take $q=601$ (a prime number). Using the definition: $M=-\partial{e_g}/\partial{\alpha}$, we calculate $M$ as a function of $\alpha$.  

In Fig.~\ref{fig:QOs1}, we present the data from this calculation for $U=4$ and different $V$'s. Note that for $U=4$, the critical point is $V_c = 0.7$, and the two inversion points are $V_{i1} = 0.96$ and $V_{i3}=2.41$. Thus, in Fig.~\ref{fig:QOs1}, by decreasing $V$, we go across the critical point from the Kondo singlet into the N\'eel phase, with inverted quasiparticle bands. This data shows that  we do get dHvA oscillations in the Kondo singlet phase close to the critical point, but they are less prominent compared to the oscillations in the AFM phase. Note that for $V > V_{i3}$, that is before the band inversion starts, we find the dHvA oscillations to be completely absent. Moreover, in the range $V_{i1} < V <V_{i3}$, we begin to see very faint signatures of the oscillations only very close to $V_{i1}$. But for the $V$'s sufficiently less than $V_{i1}$, with the bands having inverted, the magnetic quantum oscillations are clearly visible and become more pronounced upon decreasing $V$. Look at the $M$ for smaller values of $\alpha$ ($\lesssim 0.25$) in Fig.~\ref{fig:QOs1}(a), or the $M/\alpha$ for $1/\alpha \gtrsim 4$ in Fig.~\ref{fig:QOs1}(b). These numerical findings in the insulating ground state of the SPAM strongly suggest that the inversion of the charge quasiparticle bands is an important factor in realizing the dHvA oscillations in the Kondo insulators. 

In Fig.~\ref{fig:QOs2}, we present the magnetic quantum oscillation data for a fixed $V$ ($=0.6$) and different $U$'s. In this case, the system is in the N\'eel phase for $U > 2.45$, and the inversion occurs for $U > 1.59$. This data leads to the same conclusions as drawn above. That is, the amplitude of the dHvA oscillations grows by decreasing the effective Kondo coupling (by increasing $U$ here), and the inversion of the charge quasiparticle bands is necessary for these oscillations to occur. 

\begin{figure*}[!htb]
\centering
\includegraphics[width=\textwidth]{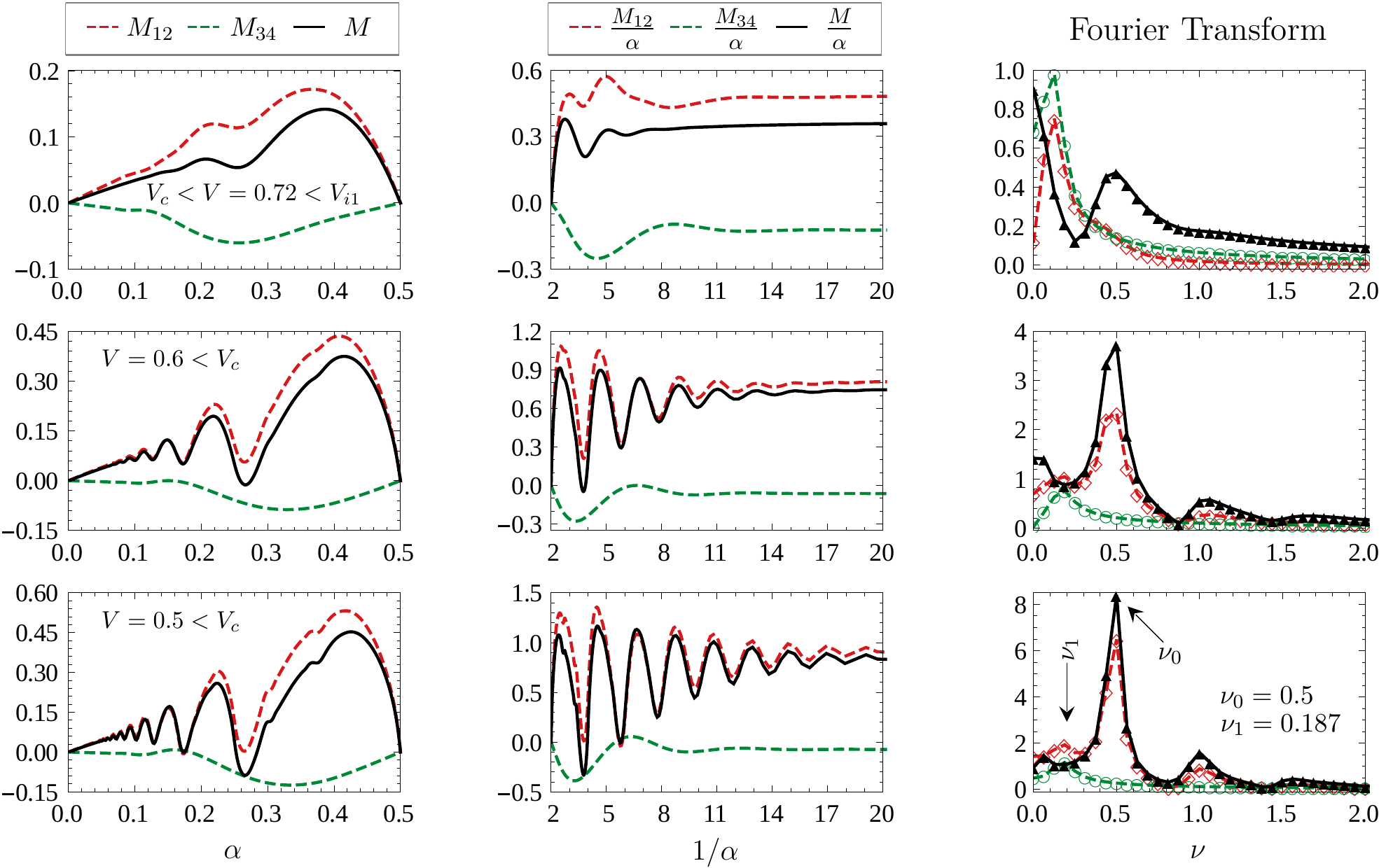} 
\caption{The magnetizations $M_{12}$ and $M_{34}$  from the two types of charge quasiparticles described by the dispersions $E_{\k,1(2)}$ and $E_{\k,3(4)}$, respectively. The total magnetization $M=M_{12}+M_{34}$. Notably the frequency $\nu_1$, with which the $M_{34}$ oscillates, is absent in the Fourier transform of the total $M$ because of a cancellation by the oscillations of $M_{12}$ with same frequency.}
\label{fig:M12-M34}
\end{figure*}

\begin{figure}[htbp]
\centering
\includegraphics[width=.45\textwidth]{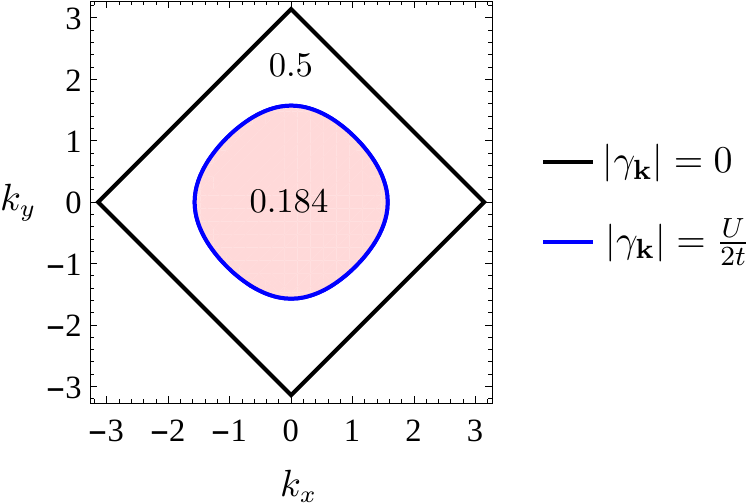}
\caption{The area of the half Brillouin zone (enclosed by the contour, $|\gamma_\k|=0$) corresponds to the oscillation frequency, $\nu_0=0.5$, of the net magnetization, $M$. The shaded region of area $0.184$ (in units of $4\pi^2/a^2$) is enclosed by the contour, $|\gamma_\k|=U/2t$, where the energy of the localized quasiparticles crosses the conduction electron quasiparticle bands in the limit $V \rightarrow 0$. Notably, the frequency, $\nu_1=0.187$ (of the oscillations of $M_{34}$, cancelled by the out of phase oscillations of $M_{12}$) compares quite well with this area.}
\label{fig:area-nu1}
\end{figure}

By doing the Fourier transformation of the $M/\alpha$ vs. $1/\alpha$ data in Figs.~\ref{fig:QOs1}(b) and~\ref{fig:QOs2}(b), we find the dominant frequency of the dHvA oscillations to be $\nu_0=0.5$, as shown in Figs.~\ref{fig:QOs1}(c) and~\ref{fig:QOs2}(c). This frequency corresponds precisely to the area of the half BZ of the square lattice (that is, the area enclosed by the contour, $|\gamma_\k|=0$; see Fig.~\ref{fig:area-nu1}). Recall the Onsager's relation, $A = (2\pi/a)^2 \nu$, between the area $A$ of an extremal orbit perpendicular to magnetic field on a constant energy surface in the $\k$-space and the frequency $\nu$ (in units of $h/ea^2$) of the dHvA oscillations. All of these findings for the SPAM are fully consistent with what we had obtained for the KLM~\cite{Ram2017}. But there is more to these findings on the dHvA oscillations in the SPAM, as described below.

\subsection{Hidden oscillations} 
Remember here we have two kinds of charge quasiparticles described by two different pairs of dispersions, $E_{\k, 1(2)}$ and $E_{\k,3(4)}$, shown in Fig.~\ref{fig:charge-disp}. They exhibit two separate inversion transitions. But the data in Figs.~\ref{fig:QOs1} and~\ref{fig:QOs2} does not say anything about their individual contributions to the magnetic quantum oscillations. To get an insight into this, we resolve the magnetization, $M$, into the magnetization, $M_{12}$, of the quasiparticles with dispersions $E_{\k, 1(2)}$ and the magnetization, $M_{34}$, of the quasiparticles with dispersions $E_{\k,3(4)}$, where $M_{12}+M_{34}=M$. We could do this resolution because the two sets of energy bands are separated by a gap~\footnote{Notably, the basic features of the charge quasiparticles, such as a non-zero charge gap, $\Delta_c$, and the two pairs of energies separated by a gap (like the narrow and broad bands in Fig.~\ref{fig:charge-disp}), survive even when $\alpha$ is non-zero.}. 

In Fig.~\ref{fig:M12-M34}, we present the data for $M_{12}$ and $M_{34}$ together with $M$ for three representative values of $V$ below the inversion point $V_{i1}$ and across the critical point $V_c$ for a fixed $U (=4)$. A careful look at the plots in the first two columns reveals that the $M_{34}$ oscillates out of phase with respect to $M_{12}$, and the total $M$ is generally dominated by $M_{12}$ (particularly so for the lower values of $V$). The most interesting aspect of this data, revealed by the Fourier transformation, is that the $M_{34}$ oscillates with a frequency, $\nu_1$, which is absent in the oscillations of the total $M$. It looks surprising, but we understand it as follows. See the Fourier transform of $M_{12}$ reveals two frequencies, $\nu_0$ $(=0.5)$ and $\nu_1$. Since $M_{12}$ oscillates out of phase with respect to $M_{34}$, the oscillations with frequency $\nu_1$ happen to cancel out completely. Thus, inspite of the non-trivial magnetic oscillations exhibited individually by the two kinds of quasiparticles, the net magnetization oscillates with a frequency of $0.5$ only. The dHvA oscillations in the SPAM is an interesting case of some hidden oscillations that cancel out. 

The frequency of these hidden oscillations is better resolved for smaller $V$'s when the oscillations are more prominent. We find $\nu_1 \approx 0.187$, which is very close to the area, $0.184$, enclosed by the contour $|\gamma_\k|=U/2t$. See the blue contour in Fig.~\ref{fig:area-nu1}. This contour is where the dispersions $E_{\k,1(2)}$ would touch $E_{\k,3(4)}$  in the limit $V\rightarrow 0$, or in other words, the energy of the $f$ electron quasiparticles crosses the band of the $c$ electron quasiparticles. Since $E_{\k,1(2)}$ are oppositely curved relative to $E_{\k,3(4)}$ near this contour, the corresponding oscillations of $M_{12}$ and $M_{34}$ (with frequency $\nu_1$) cancel each other.

%%%%%%%%%%%%%
\section{Conclusion}\label{sec:conclusion}
In this paper, we have studied the SPAM (symmetric periodic Anderson model) on the square lattice, using the theory of Kondo insulators that we first developed for the half-filled KLM (Kondo lattice model) in Ref.~\onlinecite{Ram2017}. Our approach appropriately produces the basic features of the insulating ground state of the SPAM. By decreasing $V$ for a fixed $U$, we discover two inversion transitions for the two types of charge quasiparticles. After the quasiparticle bands have suitably inverted, our calculations produce the dHvA oscillations in a uniform magnetic field. Although the two kinds of quasiparticles exhibit the magnetic oscillations of two frequencies, but due to a cancellation, only the oscillations with frequency 0.5, corresponding to the half Brillouin zone, survive. Thus, our theory produces a consistent physical picture of the inversion and magnetic quantum oscillations in the basic models of Kondo insulators, viz, the half-filled Kondo lattice model and the symmetric periodic Anderson model. 

%%%%%%%%%%%%%%%%%%%%%%%%
\begin{acknowledgments}
We acknowledge the financial support under UPE-II and DST-PURSE programs of JNU. We also acknowledge the use of HPC cluster at IUAC, and DST-FIST funded HPC cluster at SPS, JNU, for numerical calculations.
\end{acknowledgments}

%%%%%%%%%%%%%%%%%%%%%%%%%%%%%%%%%%%%%%%%%%
\bibliography{references}% Produces the bibliography via BibTeX.
\end{document}